\documentclass[preprint2]{aastex631}

\usepackage{amsmath}
\usepackage{graphicx}

\usepackage{graphicx}
\usepackage{bm}

\definecolor{ao(english)}{rgb}{0.0, 0.5, 0.0}

\begin{document}

\title{A Machine Learning Method for Monte Carlo Calculations of Radiative Processes}

\author{William Charles}
\affiliation{Physics Department and McDonnell Center for the Space Sciences, Washington University in St. Louis, St. Louis, MO 63130, USA}

\author[0000-0002-4738-1168]{Alexander Y. Chen}
\affiliation{Physics Department and McDonnell Center for the Space Sciences, Washington University in St. Louis, St. Louis, MO 63130, USA}

\begin{abstract}

Radiative processes such as synchrotron radiation and Compton scattering play an important role in astrophysics. Radiative processes are fundamentally stochastic in nature, and the best tools currently used for resolving these processes computationally are Monte Carlo (MC) methods. These methods typically draw a large number of samples from a complex distribution such as the differential cross section for electron-photon scattering, and then use these samples to compute the radiation properties such as angular distribution, spectrum, and polarization. In this work we propose a machine learning (ML) technique for 
efficient sampling from arbitrary known probability distributions that can be used to accelerate Monte Carlo calculation of radiative processes in astrophysical scenarios. In particular, we apply our technique to inverse Compton radiation and find that our ML method can be up to an order of magnitude faster than traditional methods currently in use.

\end{abstract}


\section{Introduction} \label{sec:intro}

Radiative processes play a central role in astrophysics, as electromagnetic radiation remained the only means we can use to observe the universe until the recent successes of gravitational wave and neutrino detectors in the last few years~\citep[e.g.][]{PhysRevD.109.022001, PhysRevLett.132.151001}.
Theoretical models, in order to make connection to observational data, must rigorously carry out the radiative transfer calculations at the source, as well as along the propagation path before the signals reach us. A successful example is the recent Event Horizon Telescope (EHT) simulation campaign~\citep[e.g.][]{2019ApJ...875L...1E, 2021ApJ...910L..13E} where sophisticated general-relativistic ray-tracing calculations were performed on numerical GRMHD models to produce radiation signatures that can be directly compared with the EHT observations.

In some physical systems, e.g.\ when the optical depth is order unity or above, radiation may also feedback onto the dynamics of the plasma emitting the radiation, through radiation energy loss as well as photon scattering. For some systems, the radiation may even produce $e^\pm$ pairs in the system, thus creating a channel for plasma supply. It is therefore necessary to incorporate the effect of radiative processes into first principles numerical simulations when the radiation feedback is important. In recent years, first-principles Particle-in-Cell (PIC) simulations with self-consistent radiation feedback or even pair production became possible, and they have been used to study plasma physics in extreme environments such as near black holes or neutron stars~\citep[e.g.][]{2019ApJ...877...53H,2019MNRAS.482L..60W,2021ApJ...919L...4C,2024MNRAS.527.11587}.

Due to the stochastic nature of radiative processes, they are most commonly calculated using a Monte Carlo approach, where the photon field is represented by a large number of photon packets, and their emission and scattering are individually computed using the corresponding cross sections. The central problem with modeling radiative processes is therefore the general problem of sampling from complicated probability distributions for the energy and direction of outgoing photons.
Some of the most well-known and commonly used existing methods for doing such sampling are inverse transform sampling and rejection sampling, which can involve using approximate forms of the original distribution or resorting to a large look-up table. An efficient general method for sampling from arbitrary probability distributions will be quite useful in facilitating faster calculations of radiative processes.

In recent years, there has been a significant push in the hardware domain towards tensor accelerators: chips that are physically optimized to quickly compute matrix multiplication, which is the rate limiting step in evaluating neural networks. A neural network is a powerful machine learning tool, derivatives and extensions of which have become popular in a large number of scientific fields in recent years, with widely varying applications including image processing, time-series prediction, and language modeling, among many others~\citep{ml_computer_vision, ml_rnns, ml_time_series}. A simple neural network consists of computational
units called neurons, arranged in one-dimensional layers. Each neuron accepts a linear combination of values output by previous neurons $x$, and outputs a value $\sigma(x + b)$, where the activation
function $\sigma$ can be any nonlinear monotonic function with a known
derivative, and $b$ is a scalar term called the bias. One of the simplest forms of a neural network is a dense feed-forward (or fully-connected) network~\citep{Cybenko_1989}. Such a network architecture consists of an input layer, some number of hidden layers, and an output layer, with each layer being fully connected to the layers before and after it. More complex network architectures vary in their construction, but the procedure to train and evaluate them is largely similar.

Using a neural network to sample from a probability distribution has been attempted before. \citet{horger2018} constructed a fully connected neural network and compared the kernel density estimate (KDE) of network-generated samples with the given probability density function (PDF) of the target distribution to train the model. In this work, we introduce a more straightforward sampling method based on neural networks that mimics the inverse sampling method, with the specific aim of speeding up accurate Monte Carlo calculations of radiative processes in astrophysical applications.

This
  paper is organized as follows. In Section~\ref{sec:methods}, we describe in
  detail our machine learning method for sampling from a probability
  distribution, both the univariate case and the multivariate case. In
  Section~\ref{sec:performance}, we validate the method using relatively simple
  test functions. In Section~\ref{sec:physics}, we apply this neural network
  technique to the problem of inverse Compton (IC) scattering, and compare its
  performance with standard Monte Carlo techniques. Finally, we discuss the
  potential applications of this method to general astrophysics problems in
  Section~\ref{sec:discussion}.

\section{The ML Sampling Method}\label{sec:methods}

\subsection{Univariate Case} \label{subsec:1dmethods}

A universal method for sampling from an arbitrary probability distribution described by the probability density function (PDF) $P(x)$ is inverse transform sampling, or inversion sampling. Let us assume that we have free access to a sample $u$ randomly drawn from the uniform distribution on the interval $[0,1]$. Inversion sampling requires first obtaining the cumulative density function (CDF) $C(x)$ by integrating the PDF:
\begin{equation}
    C(x) = \int_{-\infty}^{x} P(x')\ dx'.
\end{equation}
Then, a sample from the desired distribution can be obtained using the inverse of the CDF:
\begin{equation}
    x = C^{-1}(u).
\end{equation}

If the CDF of some distribution has a well defined analytic form and inverse, then it is trivial to sample from that distribution using this method. If that is not the case, then inverse transform sampling usually evaluates the function $C^{-1}$ using either a look-up table or some approximate functions that do have an analytical inverse.

A dense feed-forward neural network is able to approximate any continuous function between two spaces to an arbitrary degree of accuracy, as long as the network contains enough neurons~\citep{kratsios2020}. Therefore, a straightforward application of neural networks here is to approximate the inverse of the CDF, $C^{-1}$, which can map a uniformly distributed random variable to a sample from our desired probability distribution.

Our goal is to construct a dense neural network approximation of the inverse CDF $C^{-1}$ of an arbitrary probability distribution. The network has $N$ layers, with each layer $1\leq i\leq N$ containing $n_i$ neurons that are fully connected. At each layer of the network, an output vector $\bm{x}_{i+1}$ is calculated from an input vector $\bm{x}_{i}$, a bias vector $\bm{b}_i$, and a matrix of layer weights $\bm{W}_i$:
\begin{equation}
    \bm{x}_{i+1} = \sigma(\bm{W}_i\bm{x}_i + \bm{b}_i),
\end{equation}
where $\sigma$ is the activation function. The network accepts a random variable $u$ sampled from a uniform distribution between 0 and 1, plus an additional collection of model parameters $\bm{Z}$ that specify the probability distribution. The output of the network is simply a number $x$ that is expected to be an approximation of $C^{-1}(u)$. When applied to large batches, the network is trivially parallelizable to take in $M$ independent uniformly distributed random numbers $\{u_j\}$ and produce $M$ samples $\{x_j\}$ that follow the target probability distribution.

To train the network, we directly construct an objective function using the CDF of the target distribution, and define the loss function as:
\begin{equation}
    \label{eq:loss-1d}
    r = \mathrm{MSE}(C(x), u), 
\end{equation}
where $u$ is the uniform random variable that is part of the input to the network, and $x$ is the output of the network; MSE stands for the standard L2 mean-squared-error loss function,
\begin{equation}
    \mathrm{MSE}(\bm{a}, \bm{b}) \equiv \frac{1}{n}\sum_{j=1}^N(a_j - b_j)^2,
\end{equation}
where $\bm{a}$ and $\bm{b}$ are any two vectors of the same length $N$. The form of the loss function given in Equation~\eqref{eq:loss-1d} avoids actually computing the inverse of the function $C$ during the training step, which simplifies implementation.
Using the chain rule, we can calculate the derivative of $r$ with respect to every trainable parameter (typically the set of all weights and biases) in the network. Then, using a stochastic iterative process of gradient descent we train the network to accurately produce the desired target output over the entire input space by minimizing $r$ for a collection of training samples~\citep{kingma2017adam}.

The input space is the space created by the set of input variables to the network. One input is one set of values for the input variables. During training, every epoch we construct a collection of input samples, which is called a batch. The samples are usually constructed by simply picking random values for each input variable between the bounds prescribed by the domain of the problem. It is better to train neural networks on large batches of inputs at once to speed up convergence and decrease effects of over-fitting or forgetting~\citep{NEURIPS2019_dc6a7071}. 
The batch size that should be chosen for training a neural network depends heavily on the exact nature of the problem and the architecture of the network. In our application we typically choose batch sizes between $1,000$ and $10,000$. We also employ batch size scheduling, where we increase the batch size at later training epochs. This can sometimes reduce overall training loss.

\subsection{Multivariate Case}

The multivariate case is somewhat more complicated, since one usually cannot simply invert the cumulative distribution function of two variables or more. Considering a two-dimensional problem where the PDF is $P(x_1, x_2)$, one can in general still define a mapping between $(x_1, x_2)$ and a pair of uniform random variables $(u_1, u_2)$ from the unit square $[0,1]\times[0,1]$:
\begin{equation}
    (x_1, x_2) = D(u_1, u_2),
\end{equation}
where $D$ is a mapping from the unit square to the target sample space.

One way to construct such a mapping is to sample $x_1$ and $x_2$ sequentially, similar to a Gibbs sampler~\citep{4767596}. It is trivial to see from the definition of conditional probability
\begin{equation}
\begin{split}
    P(x_1\cap x_2) &= P(x_1|x_2)P(x_2) \\
    &= P(x_2|x_1)P(x_1),
\end{split}
\end{equation}
that it does not matter which dimension is sampled over first.

Suppose we have a two-dimensional PDF $P(x_1,x_2)$. We need to construct a CDF that has the defining property that $C(x_1,x_2)$ evaluates to the probability that a randomly chosen point drawn from the PDF with components $y_1$ and $y_2$ falls within $y_1\leq x_1, y_2\leq x_2$. Let us define a function
\begin{equation}
    A(x_1) \equiv \frac{\int_{-\infty}^{\infty}P(x_1, x_2)\ dx_2}{M_1} ,
\end{equation}
where $M_1$ is a constant chosen such that the integral
\begin{equation}
    \int_{-\infty}^{\infty}A(x_1)\ dx_1 = 1 ,
\end{equation}
so that $A(x_1)$ is a PDF. Then the CDF along the first dimension $x_1$ is
\begin{equation}
\begin{split}
    C_A(x_1) &\equiv \int_{-\infty}^{x_1} dx_1'\ A(x_1')\\
    &= \int_{-\infty}^{\infty}dx_2\int_{-\infty}^{x_1} dx_1'\ \frac{P(x_1',x_2)}{M_1}.
\end{split}
\end{equation}
So by inverting $C_1(x_1)$, we can sample $x_1$-values from the two dimensional PDF. Consider that for a given $x_1$-value, there exists a PDF 
\begin{equation}
    B(x_2) = \frac{P(x_2|x_1)}{M_2} ,
\end{equation}
where $M_2$ is a scalar value chosen so that the integral
\begin{equation}
    \int_{-\infty}^{\infty}B(x_2)\ dx_2 = 1.
\end{equation}
Then we construct a second CDF as
\begin{equation}
    C_B(x_2) \equiv \int_{-\infty}^{x_2} dx_2'\ B(x_2'),
\end{equation}
where $C_B(x_2)$ is referred to as a conditional CDF, because it will in general be a different shape depending on the value of $x_1$ chosen. For this reason the conditional CDF may also be written as $C_B(x_2|x_1)$.

By inverting both $C_A(x_1)$ and $C_B(x_2)$, we can find a mapping from the uniform distribution in two dimensions to the target PDF. A set of points $[x_1, x_2]$ drawn from the PDF $P(x_1,x_2)$ will be related to a set of points $[u_1, u_2]$ drawn from the uniform distribution over a unit square as
\begin{equation} \label{eq:2Dinv}
    [C_A(x_1), C_B(x_2|x_1)] = [u_1, u_2],
\end{equation}
and thus we can construct the mapping $D(u_1,u_2)$ as
\begin{equation}
\begin{split}
    D(u_1, u_2) &= [C_A^{-1}(u_1), C_B^{-1}(u_2|x_1)] \\
    &= [x_1, x_2].
\end{split}
\end{equation}
Following this general procedure, we can further extend the inversion sampling method to any number of dimensions.

\begin{figure}
    \centering
    \includegraphics[width=0.3\textwidth]{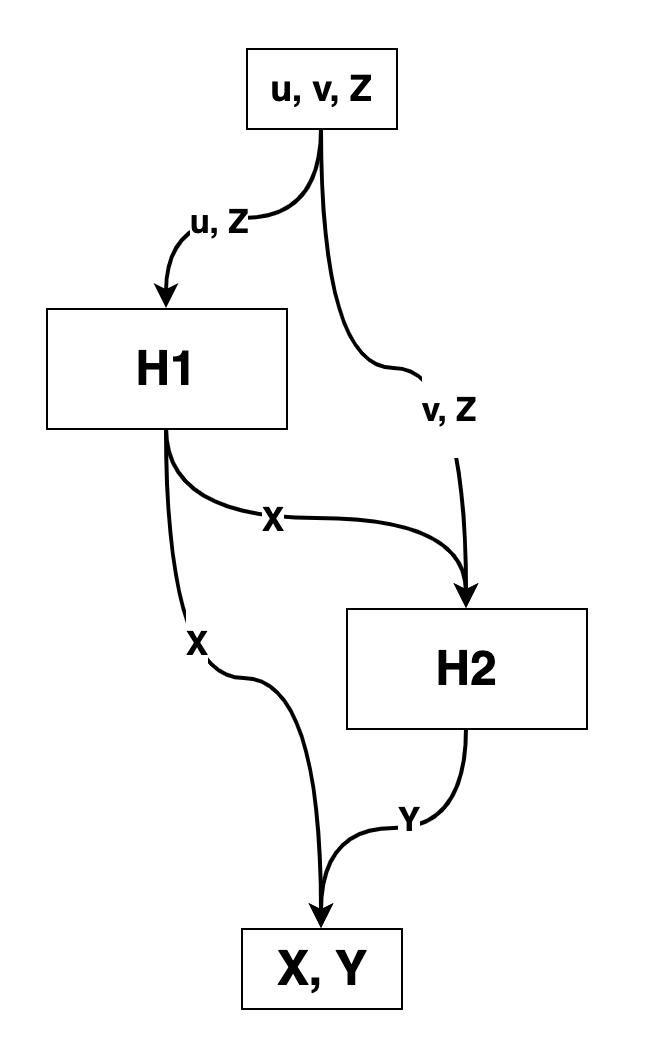}
    \caption{An overview of the Camel network architecture, so named for its two ``humps". The input to the network consists of $u$, $v$, and $\bm{Z}$, where $u$ and $v$ are sample pairs drawn from the uniform probability distribution on unit interval, $\bm{Z}$ is a set of additional parameters that the PDF may depend on, and the outputs $X, Y$ are variables drawn from the multivariate probability distribution $P(x, y, \bm{Z})$. H1 and H2 are both fully-connected neural networks.}
    \label{fig:cam}
\end{figure}

For machine learning sampling in the multivariate case, we construct a neural network of $n$ connected sub-networks, where $n$ is the sampling dimension of the target probability distribution. Each sub-network is a dense network, constructed as discussed in Section~\ref{subsec:1dmethods}.

We connect the sub-networks so that given an input of an $n$-dimensional vector of uniform random values $\bm{u}$ and a vector of auxiliary variables $\bm{Z}$ which may effect the shape of the PDF but are not sampled over, the output is a point sampled from the $n$-dimensional probability distribution as follows:
\begin{equation}
\begin{split}
    \mathrm{NN}(\bm{u}, \bm{Z}) &= \bm{x}, \\
    x_1 &= H_1(u_1, \bm{Z}), \\
    x_2 &= H_{2}(x_{1}, u_{2}, \bm{Z}), \\
    \vdots \\
    x_{n-1} &= H_{n-1}(x_{n-2}, ..., x_1, u_{n-1}, \bm{Z}), \\
    x_n &= H_n(x_{n-1}, ..., x_1, u_n, \bm{Z}),
\end{split}
\end{equation}
where each $H_{i}$ is a sub-network that approximates the inverse of the conditional CDF $C_{i}$. Figure~\ref{fig:cam} shows a schematic diagram of this architecture in the case of two-dimensional multivariate sampling.

Using the conditional CDFs along each dimension, we can directly construct an objective function for training the network. In the multivariate case the objective function is a sum over loss terms for each conditional CDF,
\begin{equation}
\begin{split}
    r = \mathrm{MSE}(C_1(x_1), u_1) + \mathrm{MSE}(C_2(x_2|x_1), u_2) \\
    + ... + \mathrm{MSE}(C_n(x_n|x_1, ..., x_{n-1}), u_n)
\end{split}
\end{equation}
where we again use the standard L2 mean-squared-error loss function. 

\section{Method Validation} \label{sec:performance}

\subsection{One-Dimensional Distribution} \label{subsec:1d}

First, we choose a one-dimensional test distribution to illustrate the simplest case of sampling using our ML method: 
\begin{equation}
    F_\theta(x) = \left[J_1(10 x \cos\theta)\cos(\theta x) + 0.6\right]\sin(\pi x),
\end{equation}
where $J_1$ is the Bessel function of the first kind of order one. In this case, $x$ is the dimension to be sampled over on the range $[0, 1]$, and $\theta$ is an auxiliary variable that controls the shape of the function. The function $F_{\theta}$ is not normalized, so we construct the PDF by dividing by the integral of $F_{\theta}$ over the domain:
\begin{equation}
    P_\theta(x) = \frac{F_\theta(x)}{M(\theta)},\quad M(\theta) = \int_0^1 F_\theta(x)\,dx.
\end{equation}
Then the CDF can be written as:
\begin{equation}
    C_\theta(x) = \int^{x}_0 P_\theta(x')\,dx',
\end{equation}
which can be readily computed numerically.

The network we use in this case is a simple feed-forward dense neural network. We built the network with four hidden layers, each with 32 neurons. We trained the network until the training loss converges to the order of $10^{-7}$. Training is done using the L2 mean squared error loss of the input and the CDF evaluated at the output. This loss is written as
\begin{equation}
    \mathrm{Loss} \equiv \frac{1}{n}\sum_i^n(u_i - C_{\theta_{i}}(\mathrm{NN}(u_i, \theta_{i})))^2
\end{equation}
where $\mathrm{NN}(u, \theta)$ is the output of the neural network evaluated with input values $u$ and $\theta$, and $n$ is the batch size. During training, $\theta$ is chosen randomly from a list of $500$ values between $0$ and $\pi$, and the batch size is 10,000 input samples. This way, during testing and evaluation we can ensure that we are testing on values of $\theta$ that haven't before been seen by the network during training.

\begin{figure}
    \centering
    \includegraphics[width=0.45\textwidth]{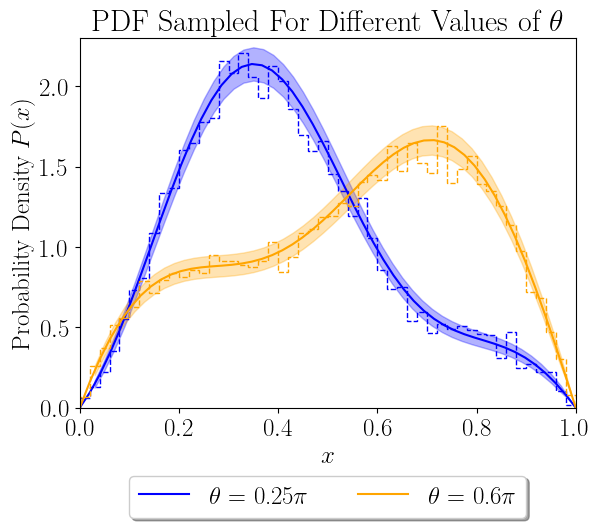}
    \caption{A histogram of $10^4$ samples generated by the neural network (dotted lines) plotted over the analytic PDF (solid lines) for different values of $\theta$, not seen by the network previously during the training process. The expected statistical noise is shown as a shaded region around the PDF.}
    \label{fig:1d_samples_err}
\end{figure}

The statistical sampling noise obtained when drawing $N$ pseudo-random samples in a numerical context grows approximately as $\frac{1}{\sqrt{n_b}}$ where $n_b$ is the number of samples in one bin of a histogram of drawn samples. The accuracy of the NN sampling method is illustrated in Figure~\ref{fig:1d_samples_err}, where we plot histograms of sampled values on top of the analytic PDFs, along with the expected statistical error as shaded regions, for some different values of $\theta$.

\subsection{Two-Dimensional Gaussian Distribution} \label{subsec:2d}

To illustrate the use of our specific network architecture in the multivariate sampling case, we also consider the basic example of a two-dimensional Gaussian distribution. We connect two sub-networks together to form the architecture shown in Figure~\ref{fig:cam}. Sampling from a multidimensional Gaussian has been thoroughly solved in the literature~\citep[e.g.][]{Box_Muller_1958}, but is included here as a proof of concept for the application of our machine learning method to a multivariate distribution.

\begin{figure}
  \centering
    \includegraphics[width=0.22\textwidth]{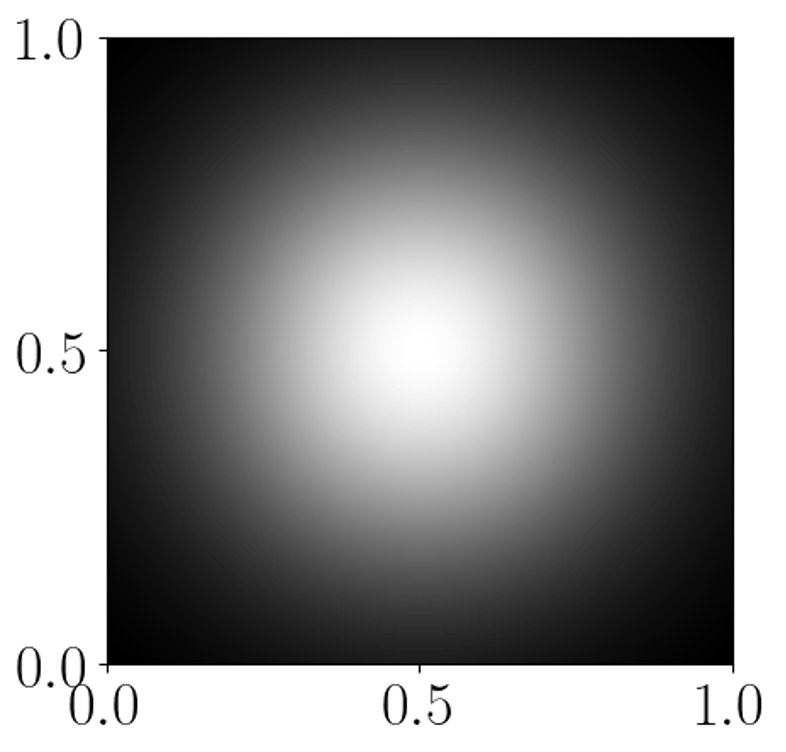}
    \includegraphics[width=0.22\textwidth]{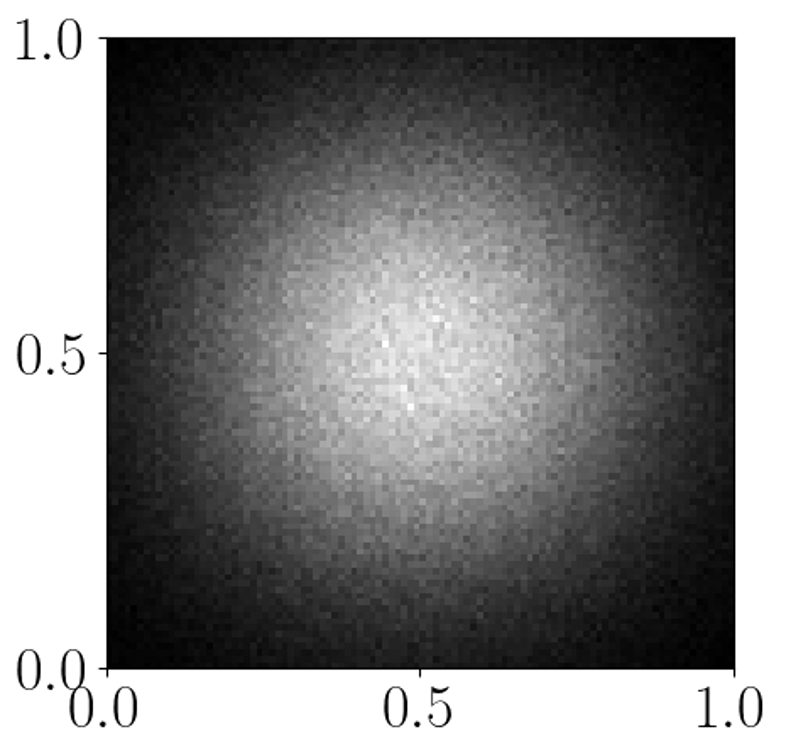}
  \caption{\emph{Left}: A standard two-dimensional Gaussian distribution. \emph{Right}: A histogram of $10^6$ samples generated by the neural network sampler after training.}
  \label{fig:2d}
\end{figure}

The PDF in this case contains no auxiliary variables that change the shape of the distribution,
\begin{equation}
    P(x,y) = e^{-8\left(\left(x-\frac{1}{2}\right)^2 + \left(y-\frac{1}{2}\right)^2\right)}.
\end{equation}
The network contains more parameters in this case. Once again we train the network until the mean squared error loss is on the order of $10^{-7}$. Figure~\ref{fig:2d} contains a side-by-side comparison of a histogram of the network output given inputs randomly sampled from the uniform distribution, and the target distribution of the two-dimensional Gaussian.

\section{Application to Inverse Compton Scattering} \label{sec:physics}

Compton scattering is a fundamental QED process that is ubiquitous in astrophysics. The limit where the moving electron participating in the scattering process has significant kinetic energy compared to the photon is called \emph{inverse Compton} (IC) scattering. In this limit, energy is usually transferred from the electron to the outgoing photon, powering high-energy radiation. For example, many active galactic nuclei (AGN) exhibit a two-humped spectrum where the high energy hump is believed to come from inverse Compton scattering of seed photons off electrons accelerated in the jet~\citep[e.g.][]{2019ARA&A..57..467B}.

In recent years, inverse Compton scattering processes have been self-consistently incorporated into first-principles kinetic simulations of pair-producing gaps~\citep[e.g.][]{Chen2020-ph,2023arXiv231012532K} and magnetic reconnection~\citep[e.g.][]{2024MNRAS.527.11587}. One of the challenges of including such radiative processes self-consistently in a fully kinetic simulation is the cost of the Monte Carlo calculations of the scattering process. Therefore, we would like to apply the machine learning sampling technique developed in Section~\ref{sec:methods} and~\ref{sec:performance} to this process and measure its performance.

\subsection{Half-Precision and Energy Mapping}

By design, our neural network will output a value between $1$ and $-1$. In order to achieve high performance, we use half-precision variables during the training and application of our network. In general this can potentially cause problems for target values very close to $1$ or $-1$ due to floating point error.
For 16-bit half-precision, the largest number less than $1$ that can be represented is around $0.9995$, and we want the network to be able to accurately resolve the scattering regime where the energy of the upscattered photon $\epsilon_\mathrm{ph}$ is very close to the minimum energy allowed by kinematic constraints, which may be on the order of $10^{-8}$. Therefore we need an algorithm to map the output interval $[-1, 1]$ to the energy space spanning many orders of magnitude.

For a given inverse Compton interaction we define the mapping between the upscattered photon energy spectrum $\epsilon_\mathrm{min} < \epsilon_\mathrm{ph} < \epsilon_\mathrm{max}$ and the range of network outputs $-1 < x_\mathrm{target} < 1$ as:

\begin{equation}
    \label{eq:xtoE}
    x_\mathrm{target} = \frac{1}{\beta}\tanh^{-1}{\left[\left(\frac{2(\epsilon_\mathrm{ph} - \epsilon_\mathrm{min})}{\epsilon_\mathrm{max} - \epsilon_\mathrm{min}} - 1\right) \frac{1}{\alpha}\right]}  ,
\end{equation}

\noindent or equivalently,

\begin{equation}
    \epsilon_\mathrm{ph} = \frac{1}{2}\left[1 + \alpha\tanh{(\beta x_\mathrm{target})}\right](\epsilon_\mathrm{max} - \epsilon_\mathrm{min}) + \epsilon_\mathrm{min} ,
\end{equation}

\noindent where $\alpha$ and $\beta$ are parameters chosen such that 

\begin{equation}
    \alpha\tanh{(\beta x)} = \begin{cases}
      1, & \text{if } x = 1, \\
      -1, & \text{if } x = -1.
    \end{cases}
\end{equation}

\noindent Thus the network can learn to produce $x_\mathrm{target}$, and by appending a few post-processing operations we recover the physical outgoing photon energy. This mapping is chosen so that the upper and lower energy limits are hard-coded into the network, and also to increase fidelity at the high and low ends of the energy spectrum by reducing the distance between distinct representable energy values. The main reason that informs this choice is that depending on parameters, IC scattering may tend to exclusively produce outgoing photons near the lowest end (Thomson regime) or highest end (Klein-Nishina regime). This mapping makes no difference to the overall explanatory power of the network, but helps the network to be more accurate in the very low and high upscattering energies. We choose a value of $\beta=5$, which fixes $\alpha=1+9.08*10^{-5}$.

\subsection{Single Photon-Electron Scattering}\label{sec:single_scattering}

We consider first the IC radiation of a single electron passing through and upscattering an isotropic field of photons. The reason of this choice was partly motivated by the application to self-consistent radiative PIC simulations such as those by~\citet{Nalewajko_2018}, \citet{2020ApJ...895..121C}, and~\citet{chand2022}.
In many astrophysical applications, the alternative problem of a photon interacting with a distribution of electrons is of greater interest, such as ray-tracing in a GRMHD simulation~\citep[e.g.][]{2021ApJ...910L..13E}. The prescription described in this section can be easily modified to apply to this scenario as well, simply by modifying the cross section used for training.

We will use the following dimensionless notation
\begin{equation}
    \gamma = \frac{E_e}{m_e c^2},\ \epsilon_0 = \frac{E_0}{m_e c^2},\ \epsilon_\mathrm{ph} = \frac{E_\mathrm{ph}}{m_e c^2},
\end{equation}
where $E_e$ is the initial energy of the incoming electron, $E_0$ is the initial energy of the interacting photon, $E_\mathrm{ph}$ is the final outgoing energy of the upscattered photon, and $m_e c^2$ is the rest mass energy of the electron.

Let us also define the energy ratio $q$ and the parameter $b$ determining the scattering regime following~\citet{Blumenthal1970}:
\begin{equation}
    b = 4 \gamma \epsilon_0,\quad q = \frac{\epsilon_\mathrm{ph}}{b(\gamma - \epsilon_\mathrm{ph})}.
\end{equation}
This definition differs slightly from that used by~\citet{Blumenthal1970} as our $\epsilon_\mathrm{ph}$ is $\gamma$ times the $E_{1}$ in that paper.
The angle-averaged scattering cross section can be used to compute the IC spectrum:
\begin{equation}
    \label{eq:ic-spectrum}
\begin{split}
    \frac{dN_\mathrm{ph}}{dt\,d\epsilon_\mathrm{ph}} &= \frac{2\pi r_0^2m_ec^3}{\epsilon_0 \gamma^2} n(\epsilon_0) d\epsilon_0 \times \left[ 2q\ln{q}\phantom{\frac{a}{b}}\right. \\
    &\left.+ (1+2q)(1-q) + \frac{1}{2}\frac{(bq)^2}{1+bq}(1-q)\right],
\end{split}
\end{equation}
where $r_0$ is the classical electron radius, and $n(\epsilon_0)\,d\epsilon_0$ is the number density of photons in the range $[\epsilon_0, \epsilon_0 + d\epsilon_0]$~\citep{Jones1968, Blumenthal1970}.
Equation~\eqref{eq:ic-spectrum} can be understood as the energy distribution of outgoing photons, and its integral is the number of photons emitted per unit time. This is the distribution we would like to sample from using the ML sampling method.

In the case of IC radiation considered here, the input space is two-dimensional but it is only a one-dimensional sampling problem. The only variable sampled over is the outgoing upscattered photon energy. However, different values for the interacting electron and photon energies drastically change the shape of the probability distribution to be sampled from. In fact, the shape of the PDF depends for the most part only on the product $\gamma\epsilon_0$, which can be seen from Equation~\eqref{eq:ic-spectrum} if we normalize $\epsilon_\mathrm{ph}$ by $\gamma$. The minimum outgoing energy of the upscattered photon, and therefore the minimum non-zero value of the PDF, is of course determined solely by $\epsilon_0$, and the maximum outgoing energy of the upscattered photon is determined by kinematic constraints and given as~\citep{Jones1968, Khangulyan2023}

\begin{equation}
    \epsilon_\mathrm{ph,max} \approx \frac{\gamma b}{1 + b} .
\end{equation}

The range of outgoing energy values for an upscattered photon will vary significantly depending on the initial energies of the interacting electron and photon. Our neural network will output a value $x$ on the interval $[-1,1]$ by design, so we want to to regularize the target energy values to the range of NN output values for efficient learning. 

With performance being the main goal, we construct a network of minimum
working size, which turns out to be a layer size of 16 neurons and only two hidden layers. This network contains 625 total parameters. The input to this network is a tuple $(u, \log_{10}(\gamma), \log_{10}(\epsilon_{0}))$ where $u$ is a random sample drawn uniformly from $[0, 1]$ and $\gamma$ and $\epsilon_{0}$ govern the shape of the probability distribution. 

We train the network using values of $\gamma$ randomly chosen from a list of logarithmically spaced values between $10$ and $10^{10}$, and $\epsilon_{0}$ similarly chosen between $10^{-10}$ and $10^{-2}$. Choosing training values from a list ensures that during testing we can check the validity of the model on parameters not seen in the training process. As before, we train the network until the loss converges upon a target value, in this case on the order of $10^{-5}$. The amount of training needed is on the order of $10^6$ epochs. On an NVIDIA GeForce RTX 3090 GPU, this amount of training typically takes a few hours.

After training, the neural network has learned a nonlinear, smooth mapping between the input and output spaces and is able to accurately interpolate between values of electron and photon energies that it saw during training, to correctly predict scattering results for electron and photon energies that were never seen during training.

\begin{figure}
    \centering
    \includegraphics[width=0.5\textwidth]{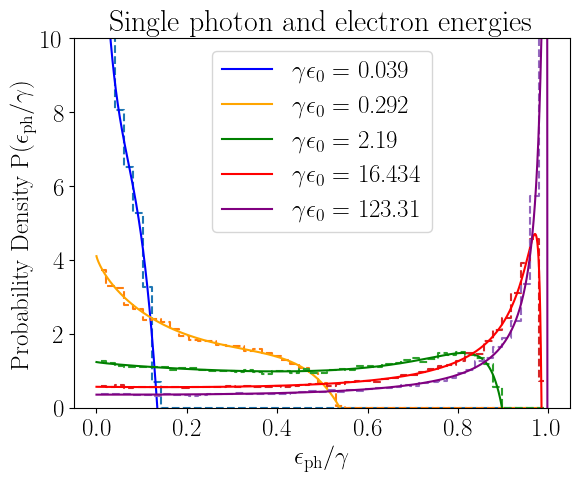}
    \caption{The transition from the Thompson to the Klein-Nishina regime for the inverse Compton scattering of an electron with energy $\gamma$ in a monochromatic isotropic photon field with energy $\epsilon_0$. The dotted lines show histograms of $10^5$ sampled energies, and the solid lines are the analytic PDF curves labelled by the product of the energies of the interacting particles. All of these energy values were not seen before by the network during training.}
    \label{fig:KN}
\end{figure}

\begin{figure*}[t]
  \centering
    \includegraphics[width=0.49\linewidth]{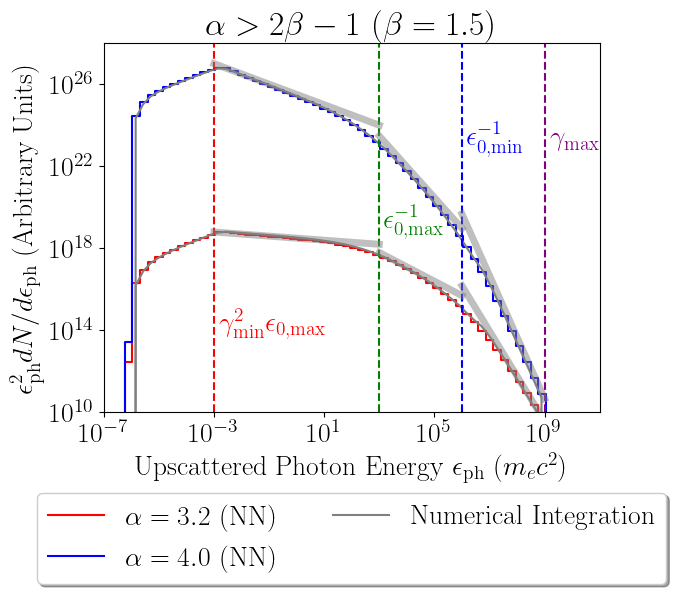}
    \includegraphics[width=0.49\linewidth]{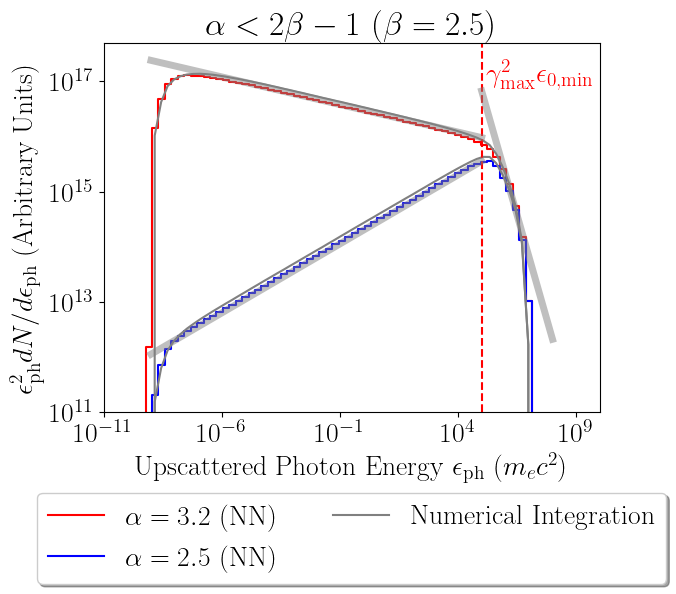}
  \caption{\emph{Left}: Numerically computed IC spectra of outgoing photon energies produced by the interaction of a target photon distribution with a power law index of $\beta = 1.5$ with electron distributions with power law indices of $\alpha = 3.2$ and $4.0$ respectively, plotted as solid black lines. Here $\gamma_{min} = 1$, $\gamma_{max} = 10^{9}$, $\epsilon_{0,min} = 10^{-6}$, and $\epsilon_{0,max} = 10^{-3}$ in dimensionless units. Histograms of sampled outgoing energy values are plotted in color, each composed of $10^6$ samples. Plotted as gray lines, we show the power law dependence that the spectrum is expected to have analytically in different energy regimes. We also mark with vertical lines the outgoing energies at which we expect to see breaks in the energy spectrum.
  \emph{Right}: A similar plot, but with photon power spectrum index $\beta = 1.5$ so that we view the regime where $\alpha < 2\beta - 1$. Here $\gamma_{min} = 1$, $\gamma_{max} = 10^{7}$, $\epsilon_{0,min} = 10^{-9}$, and $\epsilon_{0,max} = 10^{-3}$, again in dimensionless units.}
  \label{fig:dist}
\end{figure*}

We can observe closely the transition between the Thomson regime, where the outgoing photon energy distribution is peaked very close to zero, to the Klein-Nishina regime, where the outgoing energy spectrum becomes very sharply peaked near the maximum outgoing energy allowed by kinematic constraints. Even in this transition regime, shown in Figure~\ref{fig:KN} we see the network has learned to interpolate in the input space with very high accuracy.

\subsection{Electron and Photon Energy Distributions}\label{sec:distribution_scattering}

As a further test for our neural-network powered sampling method, we consider a power-law distribution of electrons interacting with a power-law background photon distribution:
\begin{equation}
    n_\mathrm{e}(\gamma) \propto \gamma^{-\alpha},\quad
    n(\epsilon_{0}) \propto \epsilon_0^{-\beta}.
\end{equation}
The power laws range from $\gamma_\mathrm{min}$ to $\gamma_\mathrm{max}$ and $\epsilon_\mathrm{0,min}$ to $\epsilon_\mathrm{0,max}$, respectively. Both $\alpha$ and $\beta$ are taken to be positive, as is most commonly the case in astrophysical scenarios.

The resulting photon spectrum is typically a broken power law, whose number of spectral breaks depends on the relative shape of the electron and target photon power-law distributions~\citep[see e.g.][]{Khangulyan2023}. In particular, when $\alpha > 2\beta - 1$, there are usually 3 spectral breaks, located roughly at energies $\gamma_\mathrm{min}^2\epsilon_\mathrm{0,max}$, $\epsilon_\mathrm{0,min}^{-1}$, and $\epsilon_\mathrm{0,max}^{-1}$. The 4 segments of the broken power law generally have distinct spectral indices:
\begin{equation}
    \label{eq:broken-power-law}
    n_\mathrm{ph} \propto \begin{cases}
      \epsilon_\mathrm{ph}^{-\beta}, & \text{if } 
      \gamma_\mathrm{min} <
      \epsilon_\mathrm{ph} < \gamma_\mathrm{min}^{2}\epsilon_\mathrm{0,max}, \\
      \epsilon_\mathrm{ph}^{-(\alpha + 1)/2}, & \text{if } \gamma_\mathrm{min}^{2}\epsilon_\mathrm{0,max} < \epsilon_\mathrm{ph} < \epsilon_\mathrm{0,max}^{-1}, \\
      \epsilon_\mathrm{ph}^{-(\alpha + 1) + \beta}, & \text{if } \epsilon_\mathrm{0,max}^{-1} < \epsilon_\mathrm{ph} < \epsilon_\mathrm{0,min}^{-1}, \\
      \epsilon_\mathrm{ph}^{-(\alpha + 1)}, & \text{if }  \epsilon_\mathrm{0,min}^{-1} < \epsilon_\mathrm{ph} < \gamma_\mathrm{max}.
    \end{cases}
\end{equation}

In the other regime where $\alpha < 2\beta - 1$, there are only 2 spectral breaks at $\epsilon_{\rm ph} = \gamma_{\rm max}^2\epsilon_{\rm 0, min}$ and $\epsilon_{\rm 0, min}^{-1}$, with spectral indices:

\begin{equation}
    \label{eq:broken-power-law}
    n_\mathrm{ph} \propto \begin{cases}
      \epsilon_\mathrm{ph}^{-(\alpha+1)/2}, & \text{if } 
      \epsilon_\mathrm{ph} < \gamma_\mathrm{max}^{2}\epsilon_\mathrm{0,min}, 
      \\
      \epsilon_\mathrm{ph}^{-\beta}, & \text{if } \gamma_\mathrm{max}^{2}\epsilon_\mathrm{0,min} < \epsilon_\mathrm{ph} < \epsilon_{\rm 0, min}^{-1}, \\
      \epsilon_\mathrm{ph}^{-(\alpha + 1)},& \text{if }\epsilon_{\rm 0, min}^{-1} < \epsilon_\mathrm{ph} < \gamma_{\rm max}
    \end{cases}
\end{equation}
Between the second and third segment, only one would be present for a given parameter regime. In the case where $\gamma_{\rm max} < \epsilon_{\rm 0, min}^{-1}$, the third segment will not be present, which is the case in our tests. Conversely, if $\epsilon_\mathrm{0,min}^{-1} < \gamma_{\rm max}^2\epsilon_\mathrm{0,min}$, or equivalently $\gamma_{\rm max} > \epsilon_{\rm 0, min}^{-1}$, then the second segment would be absent, and the spectral break is at $\epsilon_{\rm ph} = \epsilon_{\rm 0, min}^{-1}$.

Figure~\ref{fig:dist} shows the output from our trained neural network, compared to the theoretical values computed from direct numerical integration. All of the expected slopes and breaks are marked on the plot, with the histograms of outgoing energies generated by the NN sampling method plotted over the direct numerical integration of the inverse Compton scattering cross sections. The plot of each numerically integrated spectrum is normalized to have the same area under the curve as the histogram produced by the NN. 

Each spectrum plotted in Figure~\ref{fig:dist} is made by taking $10^6$ samples in total. If we take $10^6$ random samples of electron and photon energies from power law distributions over a range of $15$ orders of magnitude, we will not sample any very high energy particles. To avoid this problem we make two logarithmically spaced lists of photon energies and electron energies respectively, and then convolve those lists, taking only $100$ samples for each pair of interacting energies. Those samples are then weighted by the power law distributions of the electron and photon energies, and also by the relative total cross section per energy, and added to an overall histogram.

It can be seen in both the left and right plots of Figure~\ref{fig:dist} that there is very good agreement between the spectrum calculated by our NN sampling method, the standard numerical integration of the cross section over photon and electron energy distributions, and the characteristic features of the outgoing energy spectrum of upscattered photons that are expected analytically.

\subsection{Technical Performance}

To quantitatively analyze the accuracy of our neural network sampling method, we compare it to the inverse transform sampling method using the Jensen-Shannon (JS) divergence as a metric. The JS divergence between two probability distributions $P(x)$ and $Q(x)$ which share the same support $\chi$ is given as 

\begin{equation}
    \label{eq:JSDiv}
\begin{split}
    \mathrm{D_{JS}}(P||Q) = -\frac{1}{2}\sum_{x\in \chi}\left(P(x)\log\left(\frac{P(x)}{M(x)}\right)\right.\\\left. + Q(x)\log\left(\frac{Q(x)}{M(x)}\right)\right) ,
\end{split}
\end{equation}

\noindent where $M(x)$ is a mixture distribution

\begin{equation}
    M(x) = \frac{1}{2}(P(x) + Q(x)) .
\end{equation}

\begin{figure}
    \centering
    \includegraphics[width=\linewidth]{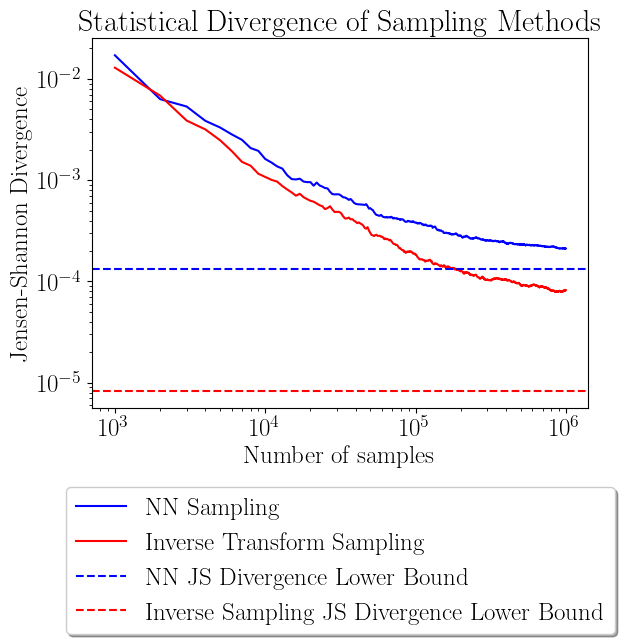}
    \caption{The Jensen-Shannon distance, the square root of the divergence, between the target cross section and samples drawn using inverse transform sampling and NN sampling for one IC radiative interaction. Here $\epsilon_0 = 10^{-4.5}$ and $\gamma = 10^5$.}
    \label{fig:jsdiv}
\end{figure}

\noindent We construct a probability distribution $P$ by choosing a specific combination of electron and photon energies in the inverse Compton regime and calculating the PDF for the upscattered photon energy. Then we draw samples from that PDF using both inverse transform sampling and our neural network sampling method. In Figure~\ref{fig:jsdiv} we plot the JS distance, which is the square root of the divergence, for each method as a function of number of samples drawn.

The neural network output is an approximation of the inverse CDF of the probability distribution, so by inverting a vector of output values and taking the derivative, we can construct the PDF approximation which has been learned by the network. The divergence between this approximation and the original PDF should provide a lower bound on the JS divergence of the histogram of NN samples, and the plot in Figure~\ref{fig:jsdiv} should trend to this value. For the particular energies $\epsilon_0 = 10^{-4.5}$ and $\gamma = 10^5$, that lower bound is about $1.3*10^{-4}$. The JS divergence of the inverse sampling method should trend to a value close to zero, which would be representative of the approximation involved in numerically integrating the PDF and inverting the CDF. The minimum possible divergence then depends on the specific numerical grid spacing chosen when integrating the PDF. For a resolution of $10^4$ grid points, the lower bound on the JS divergence is about $8.2*10^{-6}$. This is the grid spacing used in Figure~\ref{fig:jsdiv}. The lower bound will decrease with increasing numerical resolution.

As shown in Sections~\ref{sec:single_scattering} and~\ref{sec:distribution_scattering}, sample sizes of $10^6$ or less are sufficient for reproducing the inverse Compton spectra in a wide range of parameter regimes, spanning up to $15$ orders of magnitude in energy space. Therefore in most astrophysical regimes, the approximation error introduced by the neural network will be insignificant. If orders of magnitude more samples are required, then the lower bound on the JS divergence of the NN method corresponding to the network's intrinsic approximation error can be brought down to within an arbitrarily low tolerance by tuning parameters of the network or increasing the overall network size. Of course, a larger network will incur some additional evaluation expense at test time.

Evaluation of the neural network requires computing a matrix multiplication followed by evaluation of the activation function at each layer of the network. The matrix multiplication step dominates the computational cost of evaluating the network. For efficient utilization of GPU resources, and for ease of integration with existing codes, we implement evaluation of the neural network in CUDA. CUDA is a parallel computing platform developed by NVIDIA for utilization of GPUs for general computing tasks~\citep{cuda}. Pytorch, the library we used for training the network, also uses CUDA on the back end. However this comes with some overhead in Python, so for total optimization we implement our own custom evaluation of the network in CUDA directly, using the low-level warp matrix multiply-accumulate (WMMA) operations. For more details refer to the documentation from~\citet{cuda}.

In particular, we utilize the tensor cores which are built into NVIDIA GPUs. These tensor cores are designed specifically to speed up matrix multiplication for the purpose of rendering computer graphics efficiently. Use of these cores allows us to accelerate the matrix multiplication step of neural network evaluation in order to achieve very high throughput.

On an NVIDIA RTX 6000, utilizing tensor core matrix multiplication and using half-precision data for efficiency, we are able to get a throughput of up to $2.8*10^{10}$ scattering calculations per second, or about $28$ scatterings per nanosecond. This is about an order of magnitude more efficient than previous GPU-based \emph{in situ} implementation by~\citet{2020ApJ...895..121C}, which quoted a performance of about $10^9$ samples per second. Our implementation of the network evaluation in CUDA is bound by computational efficiency on our current architecture. The memory access step is an insignificant cost compared to the overall computing cost of evaluating the network. Therefore our method also scales very differently than methods that rely on look-up tables. This is significant since although modern GPUs have extremely high memory bandwidths (more than $1\,\mathrm{TB/s}$), the latency for random access on the global memory is typically quite high, up to order ${\sim}1000$ clock cycles. Therefore, non-contiguous memory access patterns such as binary search on large look-up tables are typically limited by memory latency, whereas our NN sampling method is entirely compute-bound.

For more complex sampling problems with additional auxiliary variables, our NN sampling method will scale directly with the scaling of matrix multiplication on the NVIDIA tensor cores. In general, the efficiency of evaluation only increases with increasing layer size, and therefore larger networks saturate the hardware better~\citep{cuda}. For sampling problems in higher dimensions, due to our proposed design of connected sub-networks there will also be a portion of the computing cost that scales linearly with increasing dimensionality, which is still much better than using large look-up tables for example.

The sampling throughput of $\sim 3\times 10^{10}$ scatterings per second is comparable to or exceed the state-of-art speed of PIC simulation loops, which renders the cost of real-time self-consistent radiation calculations computationally subdominant. This can potentially enable many novel applications in the realm of extreme plasma astrophysics~\citep[see e.g.][]{2019BAAS...51c.362U}. In addition, such a high-throughput sampling method can also greatly accelerate standalone Monte-Carlo calculations of radiative processes in general astrophysics applications. We share our implementation of both the network training in Pytorch and the network evaluation in CUDA, as well as our trained model at:~\url{https://github.com/will-charles/NASCaRP}. It is our hope that this tool will be useful to the community.

\section{Discussion} \label{sec:discussion}

In this paper we proposed a novel method of approximating inverse transform sampling using a neural network. The method is straightforwardly applicable to arbitrary multivariate distributions with multiple model parameters. The advantage of such a method is that it avoids large look-up tables when the distribution becomes highly complex, and modern specialized hardware makes evaluating neural networks extremely efficient.

We have presented a concrete example use case of inverse Compton scattering in the isotropic limit. Using our machine learning sampling method, we have improved the performance of such calculations by an order of magnitude. The increase in computational efficiency is especially valuable in the context of \emph{in situ} radiation calculations such as those embedded in a self-consistent PIC or MHD simulation. The fast sampling algorithm can potentially render these calculations far sub-dominant to the overall expense of the standard simulation loop, removing the overhead over regular non-radiative simulations.

A straightforward extension to our method described in this paper can be applied to the case of an anisotropic distribution of interacting photons. With more complex cross sections and more parameters, we expect the relative speed-up of our method over traditional sampling methods will further increase, as larger neural networks will more efficiently saturate the hardware, leading to higher overall sustained FLOPS. Such a model can then be applied to first-principles PIC simulations with realistic background photon field. The alternative problem of a photon going through an anisotropic plasma can be applied to both \emph{in situ} radiative transfer calculations in global MHD simulations~\citep[e.g.][]{2021ApJ...919L..20D}, or as a post-processing step to obtain the synthetic light curves, spectrta, and polarization~\citep[e.g.][]{2023ApJ...957....9W}.

Compared to a straightforward polynomial approximation of the outgoing photon distribution, the ML method presented in this paper can potentially be more accurate, and universally applicable to any problem, even when a polynomial approximation is hard to construct. The highly nonlinear nature of a neural network makes it possible to capture a wide range of feature scales, and the overall approximation error of the network can be brought down to arbitrarily low tolerances depending on the requirements for a specific application.

An additional benefit of the machine learning sampling method is the portability of the trained model. A fully trained network can be described by its weights and biases contained in a small file. This file can then be easily shared and be loaded elsewhere, skipping the training process. For common radiative processes in astrophysics, a collection of such trained models can be beneficial to the community at large.

\begin{acknowledgments}
We thank Charles Gammie, Yajie Yuan, and Nicholas Rackers for helpful discussion. This work is supported by NSF Grants AST-2308111 and DMS-2235457.
\end{acknowledgments}

\software{Pytorch \cite{pytorch}}

\bibliography{sample631}{}
\bibliographystyle{aasjournal}

\end{document}